\documentclass{appolb}
\usepackage{graphicx}

\newcommand{\msun}{\ensuremath{M_\odot}}

\newcommand{\chimera}{{\sc Chimera}}

\newcommand{\vertex}{{\sc Vertex}}
\newcommand{\agileboltztran}{{\sc Agile-Boltztran}}
\newcommand{\AB}{{\agileboltztran}}

\newcommand{\prometheusvertex}{{\sc Prometheus-Vertex}}
\newcommand{\coconutvertex}{{\sc CoCoNuT-Vertex}}

\begin{document}
\title{The Multi-Dimensional Character of Core-Collapse Supernovae%
\thanks{Presented at the XXXIV Mazurian Lakes Conference on Physics, Piaski, Poland, September 6-13, 2015}%
}

\author{
W. R. Hix$\rm^{ab}$,
E. J. Lentz$\rm^{ba}$,
S. W. Bruenn$\rm^{c}$,
A. Mezzacappa$\rm^{bd}$,\\
O. E. B. Messer$\rm^{eab}$,
E. Endeve$\rm^{fbd}$,
J. M. Blondin$\rm^{g}$,
J. A. Harris$\rm^{bh}$,\\
P. Marronetti$\rm^{i}$, 
K. N. Yakunin$\rm^{bad}$
\address{
$\rm^a$Physics Division, Oak Ridge National Laboratory, Oak Ridge, TN 37831 USA\\
$\rm^b$Dept. of Physics \& Astronomy, University of Tennessee, Knoxville, TN 37996 USA\\
$\rm^c$Dept. of Physics, Florida Atlantic University, Boca Raton, FL 33431 USA\\
$\rm^d$Joint Institute for Computational Sciences, Oak Ridge National Laboratory, Oak Ridge, TN 37831 USA\\
$\rm^e$National Center for Computational Sciences, Oak Ridge National Laboratory, Oak Ridge, TN 37831 USA\\
$\rm^f$Computer Science and Mathematics Division, Oak Ridge National Laboratory, Oak Ridge, TN 37831 USA\\
$\rm^g$Dept. of Physics, North Carolina State University,  Raleigh, NC 27695 USA\\
$\rm^h$Nuclear Science Division, Lawrence Berkeley National Laboratory, Berkeley, CA 94720 USA\\
$\rm^i$Physics Division, National Science Foundation, Arlington, VA 22207 USA}}

\maketitle
\begin{abstract}
Core-collapse supernovae, the culmination of massive stellar evolution, are spectacular astronomical events and the principle actors in the story of our elemental origins.  
Our understanding of these events, while still incomplete, centers around a neutrino-driven central engine that is highly hydrodynamically unstable. 
Increasingly sophisticated simulations reveal a shock that stalls for hundreds of milliseconds before reviving. 
Though brought back to life by neutrino heating, the development of the supernova explosion is inextricably linked to multi-dimensional fluid flows.  
In this paper, the outcomes of three-dimensional simulations that include sophisticated nuclear physics and spectral neutrino transport are juxtaposed to learn about the nature of the three dimensional fluid flow that shapes the explosion.
Comparison is also made between the results of simulations in spherical symmetry from several groups, to give ourselves confidence in the understanding derived from this juxtaposition. 
\end{abstract}
  
\section{Introduction}
A core-collapse supernova (CCSN) marks the inevitable death of a massive star (those with masses greater than roughly 8 solar masses, M$_{\odot}$) and the birth of a neutron star or black hole. 
The center of a massive star as it nears its demise is composed of iron, nickel and similar elements, the end products of stellar nucleosynthesis.  
Above this \emph{iron core} lie concentric layers of successively lighter elements, recapitulating the sequence of nuclear burning that occurred in the core.  
Unlike prior burning stages, where the ash of one stage became the fuel for its successor, no additional nuclear energy can be released by further fusion in the iron core.  
No longer can nuclear energy release stave off the inexorable attraction of gravity.  
When the iron core grows too massive to be supported by electron degeneracy pressure, the core collapses. 
This collapse continues until the stellar core reaches densities similar to those of the nucleons in a nucleus, whereupon the repulsive core of the nuclear interaction renders the stellar core incompressible, halting the collapse.  
Collision of the supersonically falling overlying layers with the stiffened core produces the \emph{bounce shock}, which drives these layers outward.  
The strength of the rebound is determined by the nuclear equation of state (EoS), as well as the structure of the progenitor star.  
For EoSs that honor laboratory limits on the nuclear compressibility, the bounce shock is not strong enough to unbind the entire envelope of the star; therefore it stalls, sapped of energy by the escape of neutrinos and nuclear dissociation.  

The failure of this \emph{prompt} supernova mechanism sets the stage for a \emph{delayed} mechanism, wherein the intense neutrino flux, which is carrying off the $10^{53}$\ erg binding energy of the newly formed proto-neutron star (PNS), heats matter above the neutrinospheres and reenergizes the shock \cite{Wils85,BeWi85}.  
Under this \emph{neutrino reheating} paradigm, the shock remains an accretion shock until sufficiently reenergized to overcome the gravity of the PNS and the ram pressure of the infalling matter, whereupon it propagates outward, heating and transmuting the overlying layers and ejecting the envelope.  
Naturally, the strength of this heating depends on the strengths of the various interactions between neutrinos and matter. 
One dimensional (1D), spherically symmetric models for this paradigm have generally failed to produce explosions because they do not deliver sufficient energy to the envelope as a result of the strict stratification imposed by spherical symmetry. 

Models that break the assumption of spherical symmetry have achieved more success by the enhancement of the neutrino luminosity due to fluid instabilities within the PNS \cite{WiMa93,KeJaMu96}, the enhanced efficiency of the neutrino heating due to large-scale convection behind the shock (see, \eg, \cite{HeBeHi94,BuHaFr95}), and the additional outward radial turbulent pressure on the shock (see, \eg, \cite{CoOt15}). 
Breaking spherical symmetry also allows the continuation of mass accretion onto the PNS, even after the explosion is well underway \cite{MaJa09, BrLeHi16}.  
PNS instabilities are driven by lepton and entropy gradients, while convection behind the shock originates from gradients in entropy that are born from the stalling of the shock and grow as the matter is heated from below. 
However, convection does not guarantee explosions \cite{JaMu96,MeCaBr98b,MeCaBr98a}.  
Simplified simulations of the stalled shock have revealed a fundamental instability of the stalled accretion shock itself to non-radial perturbations, termed the Standing Accretion Shock Instability \cite{BlMeDe03,BlMe06}.  
The SASI favors low order modes, ultimately leading to gross distortions of the shock.  
While our understanding of these events is still evolving, it is clear that the explosion centers around a neutrino-driven central engine which is highly hydrodynamically unstable. 

\section{The Limitations of Two-Dimensional Modeling of CCSN}

For an overview of the study of the core-collapse supernova mechanism, the reader is referred to recent reviews, for example, \cite{Jank12,KoTaSu12,HiLeEn14}. 
Here, we wish to focus on the recent progress in three dimensional (3D) self-consistent models and what we can learn about the limitations of more numerous two-dimensional (2D) axisymmetric models from this handful of 3D models.  

CCSN models come with a variety of approximations and parameterizations, which limit the questions that individual models can address.  
For example, much of the study of nucleosynthesis from CCSN is based on 1D models that replace the inner workings of the supernova with a kinetic energy \emph{piston} (see, \eg,\cite{WoWe95,RaHeHo02,LiCh03}) or a thermal energy \emph{bomb} (see, \eg, \cite{ThNoHa96, NaShSa98}). 
While improved models are being developed (see, \eg, \cite{UgJaMa12,PeHeFr15}), seeking to incorporate the lessons we have learned in the past two decades about the nature of these neutrino-driven, hydrodynamically unstable explosions, the relatively low computational cost of the bomb/piston models allows grids of hundreds of models to be prepared to serve as input to galactic chemical evolution studies.  
More ambitious models that include multi-dimensional fluid flow and neutrino transport are needed to explore more fundamental questions about the nature of the CCSN mechanism.  
The most ambitious models self-consistently model the matter and neutrino field all the way to the center of the neutron star, but they are by far the most costly CCSN models. 
Less accurate, but less costly, approximations to the neutrino transport have a strong appeal since they allow for improvements in other areas like increased dimensionality, resolution or model count at a constant computational cost.  
Prominent among such lower cost schemes for modeling of CCSN are \emph{leakage schemes} (dating back to \cite{BaCoKa85,vaLa81}, but seeing a recent resurgence) where the local neutrino emission rate and the opaqueness of the overlying material are used to estimate the cooling rate and hence the neutrino luminosity.   
A simplier approach is the \emph{lightbulb} approximation \cite{JaMu96}, where the neutrino transport calculation within the proto-neutron star is replaced by a prescribed neutrino luminosity at the PNS surface.  

\begin{figure}[tb]
\centerline{%
\includegraphics[width=12.5cm]{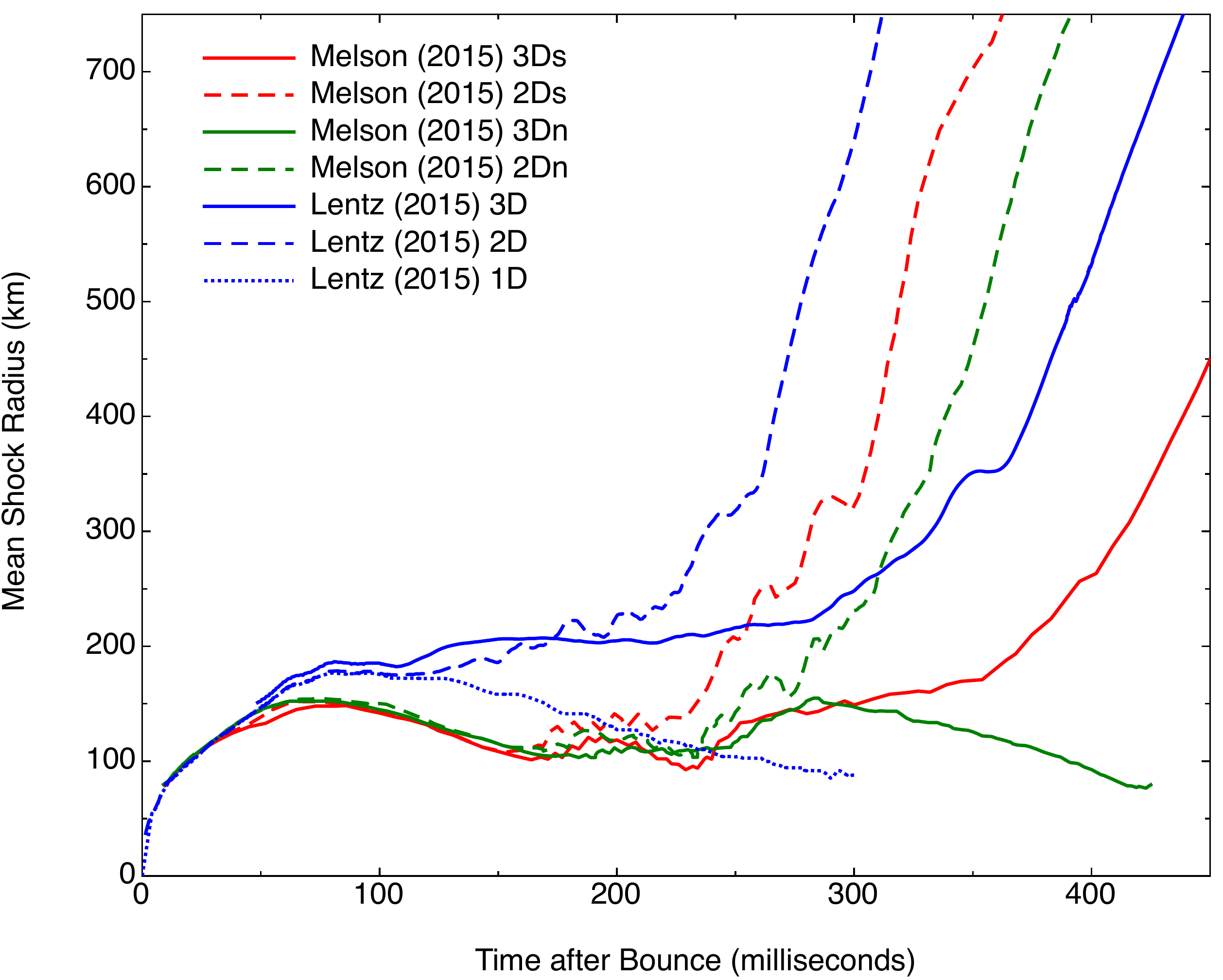}}
\caption{Comparison of the mean supernova shock radius after bounce for 3D models (solid lines) from leading neutrino transport codes using 15~\msun\ \cite{LeBrHi15} and 20~\msun\ \cite{MeJaBo15} solar metallicity progenitor stars. Results of 2D (dashed lines) and 1D (dotted lines) models are also shown.}
\label{fig:3DShock}
\end{figure}

The question of how well axisymmetric 2D models reflect the 3D reality of CCSNe has been examined in a series of simulations using variations on these simplified schemes. 
Nordhaus et al. \cite{NoBuAl10} found 3D simulations to be more favorable than 2D, producing explosions at lower neutrino luminosities, a view supported, though tempered, by Burrows et al. \cite{BuDoMu12} and Dolence et al. \cite{DoBuMu13}.  
In contrast, Hanke et al. \cite{HaMaMu12}, Couch \cite{Couc13b} and Couch \& Ott \cite{CoOt15} find 3D to be, at best, neutral compared to 2D, and likely pessimistic. 
However, ``light-bulb'' and leakage schemes do not include the complete feedback provided by self-consistent transport methods.   

\begin{figure}[tb]
\centerline{%
\includegraphics[width=12.5cm]{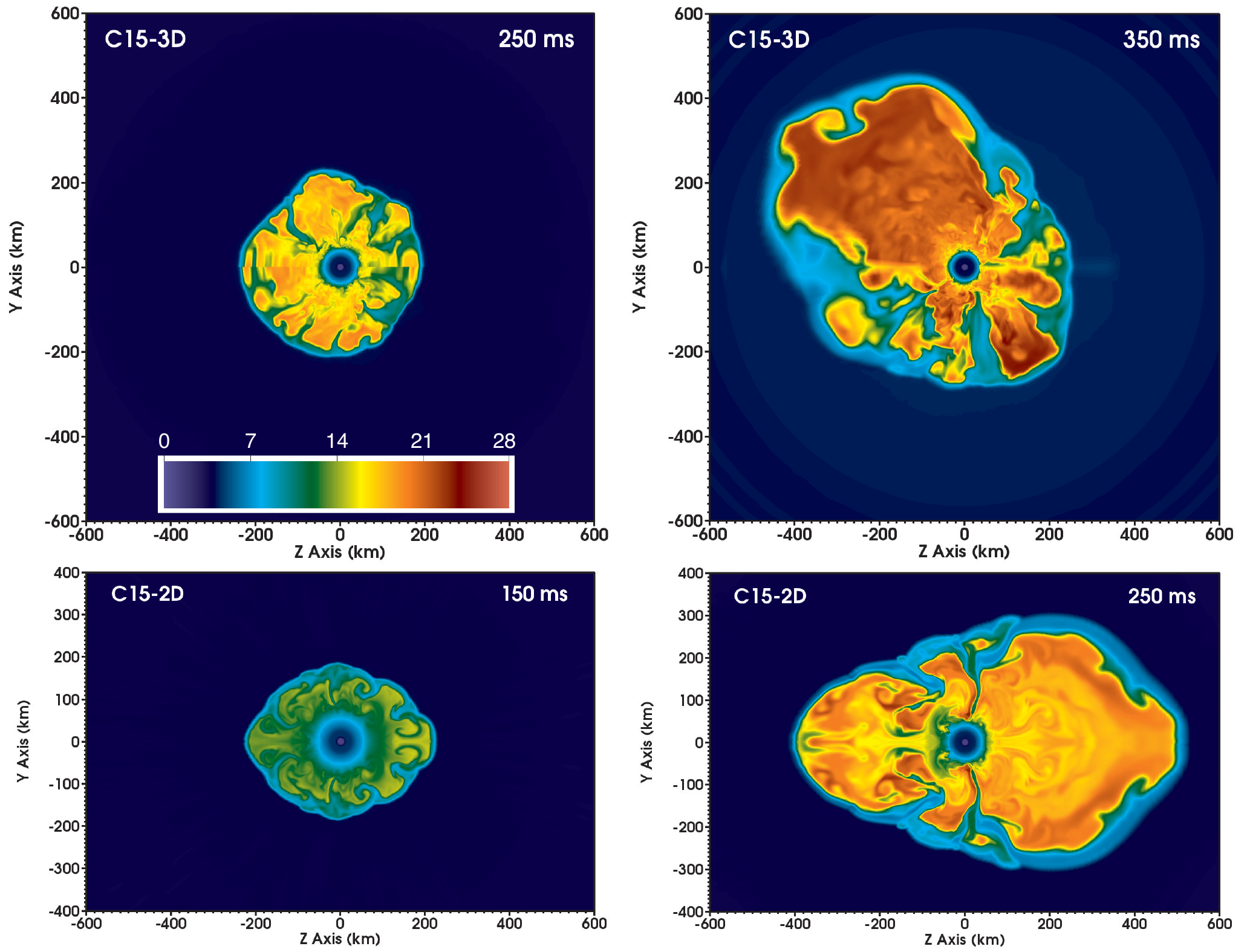}
}
\caption{The evolution of the entropy over time for 3D (top row) and 2D (bottom row) \chimera\ models of a 15 \msun\ \cite{LeBrHi15}, solar metallicity progenitor star.}
\label{fig:movie}
\end{figure}

Takiwaki et al. \cite{TaKoSu12,TaKoSu14} have shown that 3D models using the isotropic diffusion source approximation (IDSA) scheme for spectral neutrino transport, wherein the local neutrino distributions are decomposed into trapped and streaming particle components linked by a source term, also find explosions inhibited compared to their 2D counterparts.
Fryer \& Warren \cite{FrWa02,FrWa04} earlier had demonstrated that self-consistent, 3D models, albeit with \ gray (mean energy) neutrino transport, exhibit large-scale convective behavior similar to these 2D models, but that the development of explosion was delayed roughly 100 ms.
Only a handful of self-consistent 3D models using detailed spectral neutrino transport for solar metallicity stars have been published, notably a 15 \msun\ model by Lentz et al.\cite{LeBrHi15} and two variations on a 20 \msun\ model by Janka and collaborators  \cite{MeJaBo15,HaMuWo13}, both using progenitors from \cite{WoHe07}.
As shown in Fig.~\ref{fig:3DShock} (and \cite{LeBrHi15}), while 1D models reach their peak shock radius before 100 ms after bounce, the shock in two- and three-dimensional models from Lentz et al. \cite{LeBrHi15} hovers above 150 km, before progressing to explosion.  
Here too, 2D models precede their 3D counterparts to explosion by roughly 100 ms.  
The case for three dimensions delaying the explosion is even stronger for the normal physics models (2Dn/3Dn) of Melson et al. \cite{MeJaBo15}, with, as Fig.~\ref{fig:3DShock} demonstrates, 3Dn showing no sign of explosion even 150 ms after the explosion is initiated in 2Dn.  
Interestingly, the 2Ds/3Ds models of Melson et al. \cite{MeJaBo15}, where an approximate treatment of the strange-quark contributions to the nucleon spin reduces the neutrino-nucleon scattering opacity by $\sim 20\%$, show an explosion in 3D delayed by roughly 125 ms compared to 2D.  
Taken together, the models of Lentz et al. \cite{LeBrHi15} and Melson et al. \cite{MeJaBo15} show that development of explosion in self-consistent 2D models, at least for these ordinary supernovae, occurs too early by at least 100 ms.  

Another interesting similarity between these sets of models is the nature of the transition from stalled shock to explosion.  
As reflected in the mean shock radius (Fig.~\ref{fig:3DShock}), the 2D models all approach their asymptotic shock velocity within roughly 50 ms of becoming dynamic.  
In contrast, the mean shock radius in the exploding 3D models is a more gentle curve, reflecting a need for the passage of at least 100 ms for the shocks to approach their asymptotic shock velocity.  
This difference in the mean shock behavior is reflective of differences in the development of the convective engine that transfers neutrino-imparted energy to the shock.  
In Lentz et al. \cite{LeBrHi15}, where the stalled shock hovers above 150 km, the development of convection proceeds relatively rapidly, beginning around 100 ms after bounce, from a large number of small convective cells to a small number of large plumes that bridge the distance from the heating region to the shock.  
Fig.~\ref{fig:movie} illustrates this development, which hopefully, will ultimately be testable via direct observations of gravitational waves \cite{YaMeMa16} and, to a lesser extent, the neutrino signals \cite{MeDeLe16}.
The upper panels, at times of 250 and 350 ms after bounce, show the middle and late phases of this development in 3D, typified by development of a single dominant plume.  
The lower panels illustrate similar development of convection in the 2D model, by shifting to times 100 ms earlier.  
In terms of the plume scale and count, there is a great similarity between the 3D images and their 2D counterpart from 100 ms earlier.  
What is strikingly different is the entropy contrast needed to achieve this convective behavior.  
This suggests that the slower development of the explosion in 3D is the result of the need for greater neutrino heating to achieve sufficient convective development.  This prolonged heating has implications for the nucleosynthesis in these supernovae, especially the proton-richness of the ejecta \cite{HaHiCh16}.

The development of large-scale plumes is a natural consequence of the RayleighÐ-Taylor instability (see, e.g.,\cite{Shar84}), regardless of dimensionality.  
However, a fundamental difference between two- and three-dimensional fluid dynamics is the behavior of the turbulent cascade.  
In 2D, the cascade favors large scale flows, while in 3D, the cascade saps energy from these large scales in favor of smaller scales (see, e.g., \cite{DoBuMu13,EnCaBu13,Couc13b} for further discussion), effectively tearing the large eddies into smaller eddies.  
As a result, development of the large convective plumes that precede explosion in Lentz et al. \cite{LeBrHi15}, Bruenn et al. \cite{BrLeHi16}, and most successful CCSN simulations of the past two decades is artificially accelerated in two dimensions, occurring with less total neutrino heating.   
The models of Melson et al. \cite{MeJaBo15} are particularly illuminating.  
Unlike Lentz et al. \cite{LeBrHi15}, the shock in the 2D and 3D models of Melson et al. \cite{MeJaBo15} retreats to roughly 100 km by 150 ms after bounce, with little (3Dn) or no (3Ds) convective behavior reported over this time.  
Instead, the heating region is strongly SASI dominated until roughly 300 ms after bounce.  
The SASI or convective behavior of the 2D models is not discussed by Melson et al. \cite{MeJaBo15}; however, the models identified as 2Dn and 3Dn therein were extensively analyzed by Hanke \cite{Hank14}.  
Evidence for SASI activity in model 2Dn, in the form of periodicity in the time evolution of the $\ell=1$ component of the shock surface, continues to at least 260 ms after bounce, pushing the mean shock radius beyond 150 km.  
For model 2Dn, reaching such radii seems to suppress SASI activity and activate convection, leading to explosion.  
Models 3Ds and 3Dn reach similar radii somewhat later, with a similar suppressive effect on the SASI. 
However only in 3Ds, with its enhanced neutrino heating, is convection activated, leading to explosion.

\section{Comparison of methods}

\begin{figure}[tb]
\centerline{%
\includegraphics[width=12.5cm]{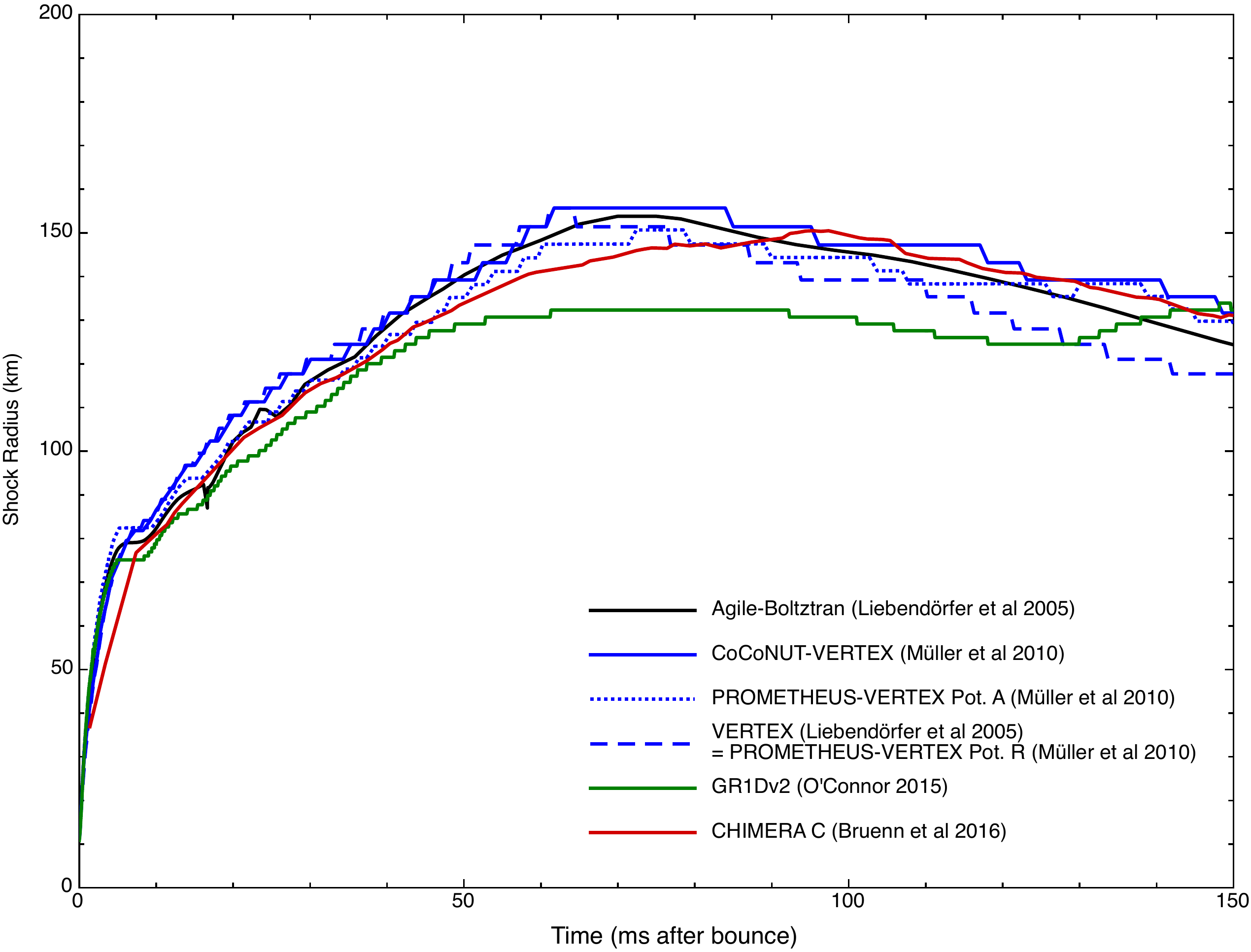}}
\caption{Comparison of the supernova shock radius after bounce as computed by leading neutrino transport codes for a 15 solar mass progenitor star \cite{WoWe95}.}
\label{fig:1Dshock}
\end{figure}

Extracting fundamental physical understanding from numerical simulations is a challenging endeavor.  
The discussion in the preceding section, as an example, draws on models from a number of groups, each using their own code, to learn about the underlying physics.  
Even among the self-consistent simulation codes, there is considerable variation in the treatment of neutrino transport, hydrodynamics, nuclear kinetics and gravity.  
Understanding the impact of these different approximations, both in the continuum limit of their analytic derivations and in their discrete implementations within the codes, requires detailed comparisons to be made between codes.  
While the necessity of such comparisons is clear to all, the work involved and the challenge of collaborating with competitors has resulted in relatively few such comparisons being published.  

In the field of core-collapse supernovae, the most influential methods comparison is that of Liebend\"orfer et al. \cite{LiRaJa05}, wherein 1D models were computed with \agileboltztran\ and \prometheusvertex, using common progenitors, a common EoS \cite{LaSw91} and common neutrino opacities \cite{Brue85}, and detailed comparisons made.  
Furthermore, the data underlying the graphs from the comparison was also made publicly available.   
This has resulted in a number of developers presenting the results of simulations utilizing the same set of progenitor and physics from Liebend\"orfer et al. \cite{LiRaJa05} with more recently developed (or improved) codes. 
\coconutvertex \cite{MuJaDi10}  and GR1D\cite{OCon15} include data from 1D simulations of this configuration and similar results for \chimera\ are forthcoming \cite{BrLeHi16b}.  
Here we provide a brief comparison of these results. 

\begin{figure}[tb]
\centerline{%
\includegraphics[width=12.5cm]{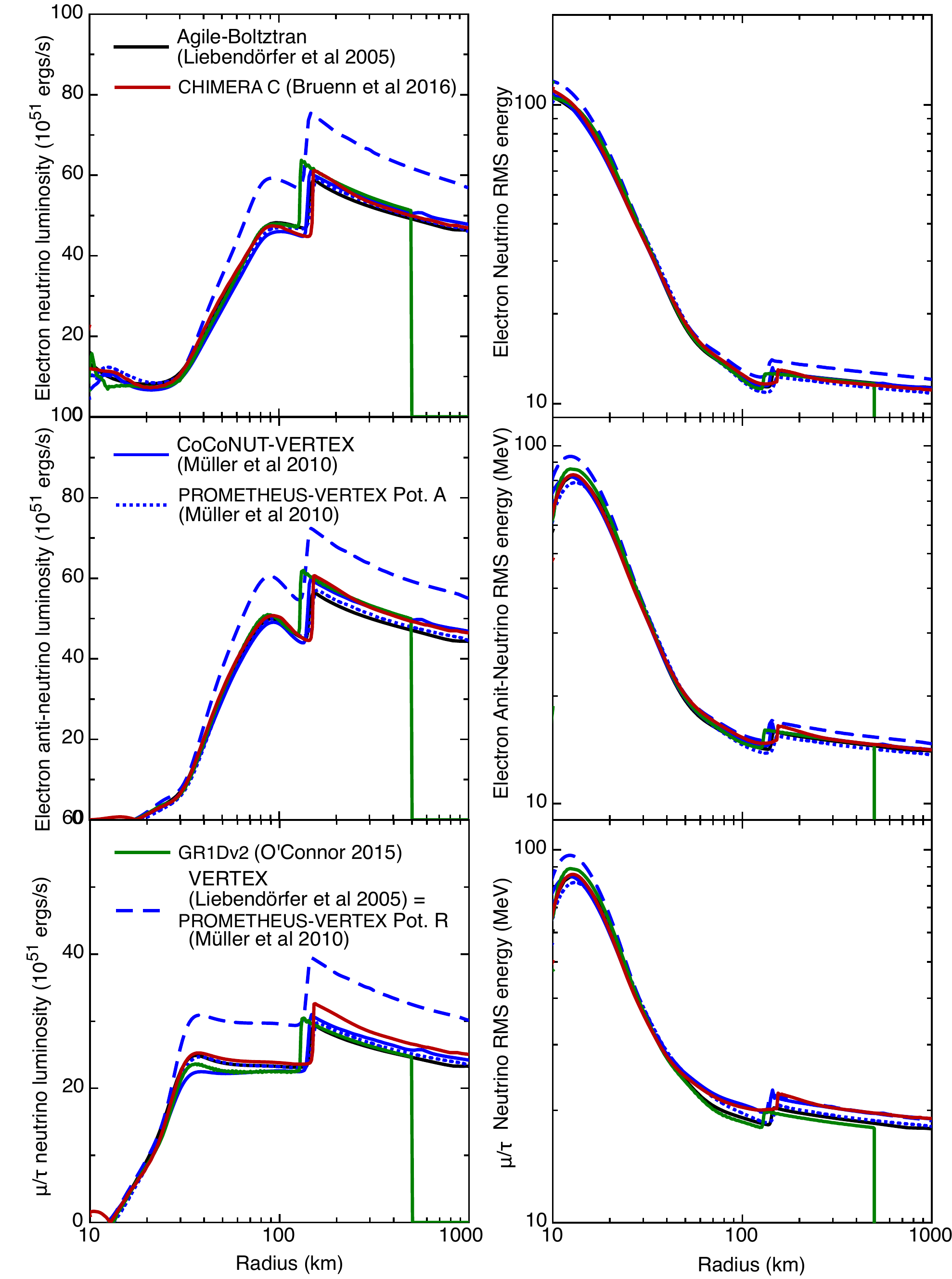}}
\caption{Comparison of Neutrino Luminosities and Root-Mean-Square (RMS) Energies 100 ms after bounce from leading neutrino transport codes \cite{LiRaJa05,MuJaDi10,OCon15,BrLeHi16} for a 15 solar mass progenitor star \cite{WoWe95}.}
\label{fig:nucomp}
\end{figure}

Fig.~\ref{fig:1Dshock} displays the radial progress of the supernova shock over 150 ms, starting from the formation of the neutron star.  
Results from 6 models are displayed.  
The black line in this figure, as well as in Fig.~\ref{fig:nucomp}, is data from Liebend\"orfer et al. \cite{LiRaJa05} for the \agileboltztran\ code.  
With its implementation of full (1D) general relativity (GR) and discrete ordinates Boltzmann neutrino transport, \AB\ has been used as a standard to which other codes are compared for more than a decade.  
The green line is data from the GR1D code \cite{OCon15}.  
The blue lines are from three variants of the \vertex\ neutrino transport code coupled to different hydrodynamic and gravitational solvers, as discussed in M\"uller et al. \cite{MuJaDi10}.  
The solid blue line is data from a model using \coconutvertex \cite{MuJaDi10}.  
The dotted blue line represents a model run with \prometheusvertex\ using the approximate GR potential A, while the dashed blue line uses \prometheusvertex\ and the older approximate GR potential R.  
Differences between Potential A and R are discussed in \cite{MaDiJa06}.  
The \prometheusvertex\ model with Potential R is included in this comparison solely for historical reasons as it is the model labeled \vertex\ in Liebend\"orfer et al. \cite{LiRaJa05}.  
Recent \prometheusvertex, most notably the models of Melson et al. \cite{MeJaBo15} discussed in the previous section, utilize Potential A.  
The red line represent a model from the C series of \chimera, the same code utilized by Lentz et al. \cite{LeBrHi15}.  
The numerical methods for the C series \chimera\ models, and the differences between the C series and the B series \cite{BrMeHi13,BrLeHi16}, are discussed in \cite{BrLeHi16b}.  
The stepped behavior in the \vertex\ and GR1D models results from their reporting of the shock position at the zone centers of their fixed radial grids, which \AB\ and \chimera\ have moving radial grids and report interpolated shock positions.

The shock comparison is limited to 150 ms for two reasons.  
First, as can be seen in Fig.~\ref{fig:3DShock}, while the mean shock radius for 2D and 3D models agrees with that of 1D models for perhaps the first 100 ms, multi-dimensional effects are clearly important beyond this point. 
Hence comparisons between spherically symmetric models gradually lose their utility beyond 100 ms. 
Second, one weakness of \agileboltztran\ compared to the other codes considered here is its treatment of the regions where matter is not in nuclear statistical equilibrium.  
In Liebend\"orfer et al. \cite{LiRaJa05}, Agile's adaptive mesh redistribution was not instrumented to honor the contact discontinuities present at the stellar compositional interfaces.  
As a result, the sharp features in entropy, electron fraction and density present in the progenitor, and maintained by the other codes, are blurred by \agileboltztran, making it a less reliable comparison standard once the matter that originated in the star's silicon layer reaches the shock, at roughly 150 ms after bounce in this model.  
The second peak seen in the GR1D shock trajectory starting 130 ms after bounce is the result of this layer, with its lower density, and hence ram pressure, reaching the shock, despite the significantly smaller shock radius of the GR1D model at this time.  
For the \chimera\ and \vertex\ models, similar features appear beyond 150 ms, while the shock trajectory for \AB\ is smooth due to mixing by the moving grid.  

The shock progress for this spherically symmetric model is quite similar among the 4 codes presented.   
\coconutvertex\ (and the older version of \prometheusvertex\ with Potential R) leads \AB\ slightly in the first 50 ms after bounce, while \prometheusvertex\ with Potential R, \chimera\ and especially GR1D, trail slightly.  
The trailing models reach their peak shock excursion somewhat later, the last being \chimera, which peaks around 90 ms after bounce, roughly 20 ms after \AB.  
The agreement between \chimera, \coconutvertex, \prometheusvertex\ with Potential A and \AB\ over the interval between 90 and 150 ms after bounce is very good, less than 10 km or roughly 2 zones in the coarsest resolved of the models.  
This period from 100-150 ms after the peak of the shock is the most important, as here, at least in \chimera\ models, the multi-dimensional models depart from their spherically symmetric complements.  

The outbreak of convection, which causes this departure, is the result of neutrino heating, so it is also important to compare the neutrino distributions in these models.  
Fig.~\ref{fig:nucomp} displays the neutrino luminosities and RMS energies at 100 ms after bounce for the 3 neutrino species considered by all of these codes.  
The agreement between the shock trajectories carries over to the neutrino quantities, with the most noticeable differences being due to the different locations of the shock.  
Accounting for shock location, the disagreement in luminosities in Fig.~\ref{fig:nucomp} between \chimera, \coconutvertex, \prometheusvertex\ with Potential A, GR1D, and \AB\ is generally less than 5\%.  
The RMS energies in Fig.~\ref{fig:nucomp} are also very similar, if one accounts for the shock location.  
The sole exception is the $\mu/\tau$ RMS energies, where \chimera\ and \coconutvertex\ are perhaps 2 MeV harder than \AB\ ($\sim 10\%$), while \prometheusvertex\ with Potential A and GR1D agree well with \AB.

%
%

\section{Summary}

The agreement seen in models from \chimera\ and the several versions of \vertex\ presented in the previous section gives us confidence that the codes can behave very similarly when using the same (albeit reduced) physics and same (albeit older) progenitor in spherical symmetry. 
This gives us confidence that the differences discussed in the first section are due to a combination of 1) multi-dimensional effects, 2) additional physics beyond the Bruenn \cite{Brue85} standard and 3) the different progenitors.  
While these provide considerable latitude for quantitative differences, the qualitative behavior seen in the models of Lentz et al. \cite{LeBrHi15} and Melson et al. \cite{MeJaBo15}, taken together, answer the question of how two-dimensional, axisymmetric simulations of CCSN differ from their more realistic three-dimensional counterparts.  
Two-dimensional models transition from stalled shock to explosion 100 ms earlier than their three-dimensional equivalents and this transition occurs much more abruptly.  
The fact that these conclusions, at least the timing of the transition, are confirmed by the earlier self-consistent (although utilizing gray neutrino transport) simulations of Fryer \& Warren \cite{FrWa04} lends credence that this conclusion will stand for all self-consistent models of the explosion in core-collapse supernovae.
A critical event leading to a successful explosion seems to be achieving a mean shock radius of 150-200 km, either directly as in Lentz et al. \cite{LeBrHi15} and Fryer \& Warren \cite{FrWa04}, or through the activity of the Standing Accretion Shock Instability as in Melson et al. \cite{MeJaBo15}, though the value of this critical radius is no doubt dependent on progenitor structure and included physics.

Acknowledgements: We thank M. Liebend\"orfer and E. O'Connor for making data from \cite{LiRaJa05} and \cite{OCon15} publicly available, and B. M\"uller for providing the data from \cite{MuJaDi10}.
This research was supported by the U.S. Department of Energy Office of Nuclear Physics and the NASA Astrophysics Theory Program (NNH11AQ72I). 
PM is supported by the National Science Foundation through its employee IR/D program. The opinions and conclusions expressed herein are those of the authors and do not represent the National Science Foundation.
The simulations here were performed via NSF TeraGrid resources provided by the National Institute for Computational Sciences under grant number TG-MCA08X010; resources of the National Energy Research Scientific Computing Center, supported by the U.S. DOE Office of Science under Contract No. DE-AC02-05CH11231; and an award of computer time from the INCITE program at the Oak Ridge Leadership Computing Facility, supported by the  U.S. DOE Office of Science under Contract No. DE-AC05-00OR22725.


\end{document}